\documentclass{nature}
\usepackage{amsmath}
\usepackage[dvipsnames]{xcolor}
\usepackage[euler]{textgreek}
\usepackage{multirow}
\usepackage{numprint}
\usepackage{pdfpages}
\usepackage{hyperref}

\newcommand{\ssr}[1]{{\textbf{\textcolor{Cyan}{SS: #1}}}}
\renewcommand{\ssr}[1]{}

\usepackage[labelfont=bf]{caption}
\usepackage[noabbrev,capitalize]{cleveref}
\usepackage{graphicx}
\usepackage{pdfpages}

\title{Universal Battery Performance and Degradation Model for Electric Aircraft}

\author{Alexander Bills$^1$, Shashank Sripad$^1$, William L. Fredericks$^1$, Matthew Guttenberg$^1$, Devin Charles$^2$, Evan Frank$^2$,  Venkatasubramanian Viswanathan$^{1,3}$}

\usepackage{graphicx}
\makeatletter
\let\saved@includegraphics\includegraphics
\AtBeginDocument{\let\includegraphics\saved@includegraphics}
\makeatother

\begin{document}

\maketitle

\begin{affiliations}
 \item Department of Mechanical Engineering, Carnegie Mellon University, Pittsburgh, Pennsylvania, 15213, USA.
 \item A$^{\wedge}$3 by Airbus, Santa Clara, California, 95050, USA.
 \item Wilton E. Scott Institute for Energy Innovation, Carnegie Mellon University, Pittsburgh, Pennsylvania, 15213, USA.
\end{affiliations}
\newpage 
\begin{abstract}
Development of Urban Air Mobility (UAM) concepts has been primarily focused on electric vertical takeoff and landing aircraft (eVTOLs), small aircraft which can land and takeoff vertically, and which are powered by rechargeable (typically lithium-ion) batteries. Design, analysis, and operation of eVTOLs requires fast and accurate prediction of Li-ion battery performance throughout the lifetime of the battery. eVTOL battery performance modeling must be particularly accurate at high discharge rates to ensure accurate simulation of the high power takeoff and landing portions of the flight. In this work, we generate a battery performance and thermal behavior dataset specific to eVTOL duty cycles. We use this dataset to develop a battery performance and degradation model (\textit{Cellfit}) which employs physics-informed machine learning in the form of Universal Ordinary Differential Equations (U-ODE's) combined with an electrochemical cell model and degradation models which include solid electrolyte interphase (SEI) growth, lithium plating, and charge loss. We show that \textit{Cellfit} with U-ODE's is better able to predict battery degradation than a mechanistic battery degradation model. We show that the improved accuracy of the degradation model improves the accuracy of the performance model. We believe that \textit{Cellfit} will prove to be a valuable tool for eVTOL designers.
\end{abstract}

\section*{Introduction}
The improving performance, increasing cycle life, and decreasing cost of lithium ion batteries spurred by the mass market adoption of personal electronics and electric vehicles (EV) has recently enabled the development of electric aircraft. \cite{Schmuch2018,Harlow2019,Tomaszewska2019,harborair,lilium} Electric aircraft convert energy stored in on-board batteries to propulsive thrust through a high voltage distribution bus, an electric motor, and a power inverter.  This mode of propulsive energy transfer eliminates the complexity typically associated with gearboxes or mechanical transmissions, affords relatively low unit costs of propulsors (inverter, motor, propeller), and increases overall powertrain efficiency compared to internal combustion or turbine engine systems.\cite{Duffy2017} This paradigm shift in the aircraft propulsion system has enabled a vast array of new hybrid and electric aircraft configurations.\cite{vtolconcepts} Many electric aircraft designs utilize distributed electric propulsion to realize novel configurations which can achieve a significant safety and efficiency advantage over conventional single or multi-engine aircraft.\cite{Kim2018} Notably, the developments in battery technology and distributed electric propulsion have opened the door to urban air mobility (UAM) by enabling the development of electric vertical takeoff and landing (eVTOL) aircraft.  As outlined by the National Aeronautics and Space Administration (NASA), UAM aims to safely and efficiently transport passengers and cargo in an urban area.\cite{2017NASAurbanair} UAM has obvious benefits of convenience and speed for mobility in some markets\cite{Antcliff2016}, and may also have an energy-efficiency advantage over ground transport\cite{langfordbridge}. 

The design trade-offs that arise from using Li-ion based batteries for electric aircraft designs are distinct from those with combustion engines, in large part due to the orders of magnitude difference in specific energy between Li-ion batteries and jet-fuel\cite{fredericks2018performance}. Compared to terrestrial electric vehicles, the performance of aircraft is much more sensitive to battery weight \cite{Raymer, Sripad2017}.  Thus, electric aircraft require careful integration and use of Li-ion battery systems. \cite{Schfer2018,fredericks2018performance} Most eVTOL aircraft are designed for the critical case: cruise to maximum range into a headwind, followed by a redirect reserve segment and a subsequent contingency landing such as a single propulsor failure. This mission profile is especially challenging to achieve close to the battery retirement state-of-health (SOH, retirement SOH is typically 80-85\%), as maximum power output is demanded at a low state-of-charge (SOC). To ensure that the co-design of electric propulsion sub-systems is consistent with the sized vehicle geometry and weights, a rapid battery performance estimation method is required in the sizing loop.\cite{sridharan2020} Importantly, the model must be supplemented with a degradation model to account for changes in performance over the lifetime of the battery.

Estimating the state-of-charge over a duty cycle and the state-of-health over the lifetime of a cell has historically been performed using physics based models\cite{Reniers2019, Ramadass2004} or empirical and data-driven machine learning models\cite{Ng2020, Petit2016, Zhang2019}.  While physical models are typically interpretable and accurate, they can often be computationally expensive.\cite{Ramadesigan_2012} On the other hand, empirical and machine learning approaches trade interpretability for speed while retaining accuracy\cite{Ng2020}.The physics informed machine learning approach can help break this trade-off. In fields such as atomistic simulations\cite{Zuo2020} and fluid mechanics\cite{Raissi2020}, encoding physical principles for data efficiency and extrapolation in machine learning methods such as neural networks has shown promising results. Machine learning has been used extensively in the field of energy storage, including being used to optimize charging protocols with closed loop and reinforcement learning based methods \cite{attia2020closed,Park2020}, to predict the cycle life of batteries based on early cycle features \cite{Severson2019}, and to estimate and forecast a battery's state of health \cite{RICHARDSON2017209,Fermin-Cueto2020}. However, there have been few works which closely integrate electrochemical battery models and constraints with machine learning methods to improve performance predictions over the full life of a battery. \cite{Ng2020} 

In this work, we develop \textit{Cellfit}, a physics-informed machine-learning model for battery performance, thermal response, and degradation. We start with a simplified electrochemical-thermal cell model which can quickly predict the performance of the cell for a duty cycle. We then add a mechanistic degradation model which includes components for modeling SEI growth, lithium plating, and active material loss. We use a physics-informed machine learning approach to improve the accuracy of the degradation model, and then we show that the the machine learning based degradation model improves the accuracy of the performance model. The model is open-source and built using the Julia programming language,\cite{bezanson2012julia}, using the DifferentialEquations.jl package\cite{rackauckas2017differentialequations}.

\begin{figure}
    \centering
        \includegraphics[width=\textwidth]{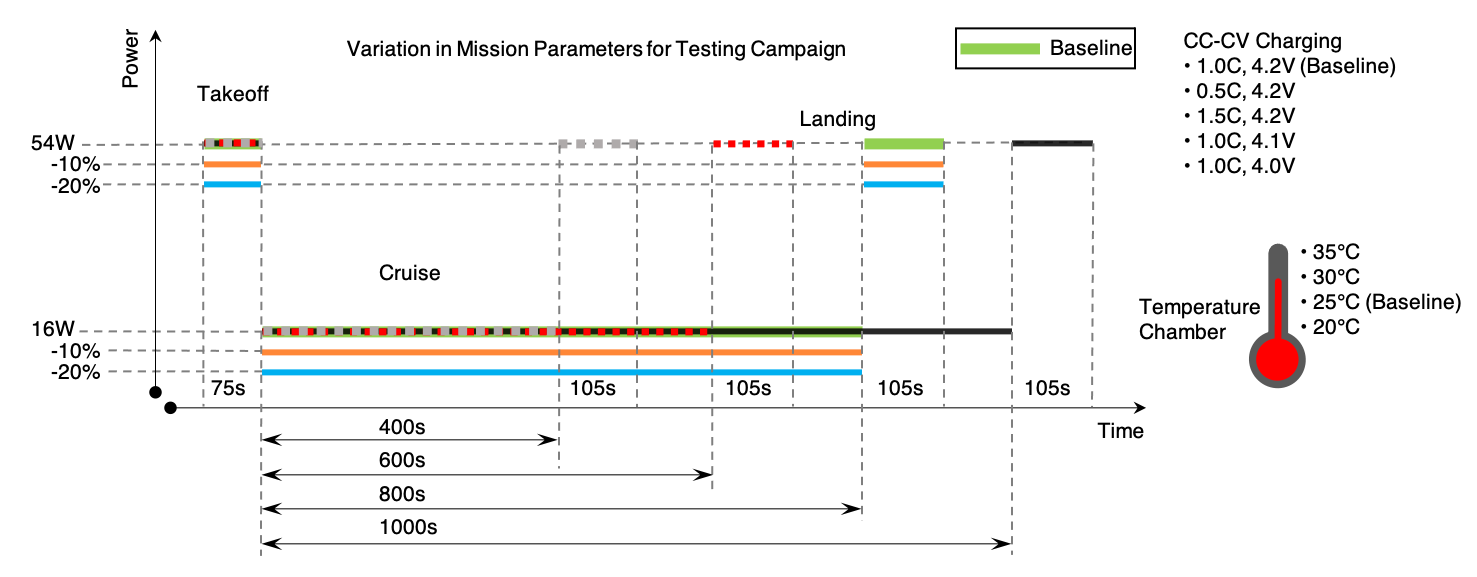}
        \caption{\textbf{eVTOL Power Profiles}  Schematic of the various mission parameters used for the cell testing campaign.}
        \label{figure1}
\end{figure}

\section*{eVTOL Battery Dataset Generation}
The development of machine learning based models requires relevant battery performance datasets to train and test the models. While datasets exist for testing Li-ion performance under a few duty cycles, \cite{saha2007battery,Harlow2019,Severson2019} the limited number of openly-available datasets remains a challenge to bring advanced machine-learning methods to battery performance prediction \cite{Severson2019,Howey2019}. \ssr{make sure to add a bunch of things from the battery genome paper} Additionally, there are no openly accessible datasets which follow the characteristic eVTOL duty cycle (figure 1).  To fill this gap, we generate an experimental battery performance dataset specific to the power requirements of an eVTOL. The dataset consists of 22 cells which sample a variety of operating conditions an eVTOL might experience during normal operation. The mission profiles follow the same generic format in all cases: (i) \textbf{Take-off:} the cell is discharged at a high constant power for a period, $t_{to}$ (ii) \textbf{Cruise:} the cell is discharged at a lower constant power for a longer duration, $t_{cr}$ (iii) \textbf{Landing:} the cell is discharged at high constant power (same rate as takeoff) for a slightly longer period of time, $t_{la}$, (iv) \textbf{Rest:} the cell is allowed to rest until it cooled to a temperature of less than 27$^\circ$C, (v) \textbf{Charging:} the cell is charged using a constant current-constant voltage (CC-CV) charging protocol, (vi) \textbf{Rest:} The cell is allowed to rest until cell temperature reached 35$^\circ$C, then allowed to rest 15 minutes further before beginning the next cycle. Every 50 cycles, each cell undergoes a capacity test wherein it is discharged at constant current until its voltage was less than 2.5V. The cells were cycled until they reached less than 2.5V or a temperature of greater than 70$^\circ$C during a typical discharge (mission profile) cycle. During each test, temperature, voltage, and current along with discharge capacity and charge capacity are recorded. The baseline profile is parameterized to test different conditions which could be encountered on a mission. Specifically, the discharge power during flight, charge current, charge voltage, ambient temperature, and cruise length are varied in a one-at-a-time fashion. The cells are divided into a training set and test sets to train and test the models (see Methods for details). The test set is further divided into extrapolative tests (conditions outside those encountered in the training set) and interpolative tests (conditions similar to those in the training sets) to facilitate further analysis of model performance.

\section*{Performance Modeling and Parameter Estimation}
For performance modeling, we use a physics inspired model\cite{bole2014adaptation} that obeys mass and charge conservation laws. The Redlich-Kister polynomial approximation is used to obtain a functional form for the equilibrium voltage of the cell. The equilibrium voltage is forced to be monotonically increasing with state of charge, thus forcing compliance with the second law of thermodynamics\cite{Bazant2013}. A diffusion-like equation is used to model transport between and within the electrodes, and conservation of charge is enforced within this equation. Electrochemical reaction kinetics is modeled with the Butler-Volmer equation. The parameters used in this model cannot be directly obtained from physical measurements. However, they are physically motivated and correspond to processes which include electrochemical reactions, reaction kinetics, and transport. This combination strikes a compromise between accuracy, computational speed, and interpretability of the model. More details of the performance model can be found in methods.

Parameter estimation for the performance model is performed against the training portion of the experementally generated eVTOL battery performance dataset. Specifically, the first cycle of each cell is used. A subset of the parameters of the performance model is chosen to capture the evolution of the performance model as the battery ages\cite{Daigle2016}, we refer to this subset as the aging parameters. The parameters of the performance model that do not evolve as the battery ages will be referred to as the performance parameters. A third set of parameters that governs how the aging parameters change over time will be introduced later. 

Parameters of battery models are often difficult to identify using local optimization methods\cite{Forman2012}. Furthermore, the dimensionality of both the performance parameters and the aging parameters makes using global optimization tractable for these sets \cite{Boyd2009}. For the performance parameters, we chose simulated annealing \cite{Kirkpatrick671} due to its ability to overcome local minima, computational feasibility in intermediate dimensionalities (18 parameters), and empirically observed good performance in this study. For the aging parameters, we chose an adaptive grid based method due to its high accuracy and because the low dimensionality of the aging parameters makes it tractable. \cite{KURCHIN2019161} The loss function used to estimate the parameters of the model is tuned to emphasize the quantities most important to the design of eVTOLs, namely, voltage error, thermal error, and maximum temperature (details can be found in methods). 

\begin{figure}
    \centering
        \includegraphics[width=\textwidth]{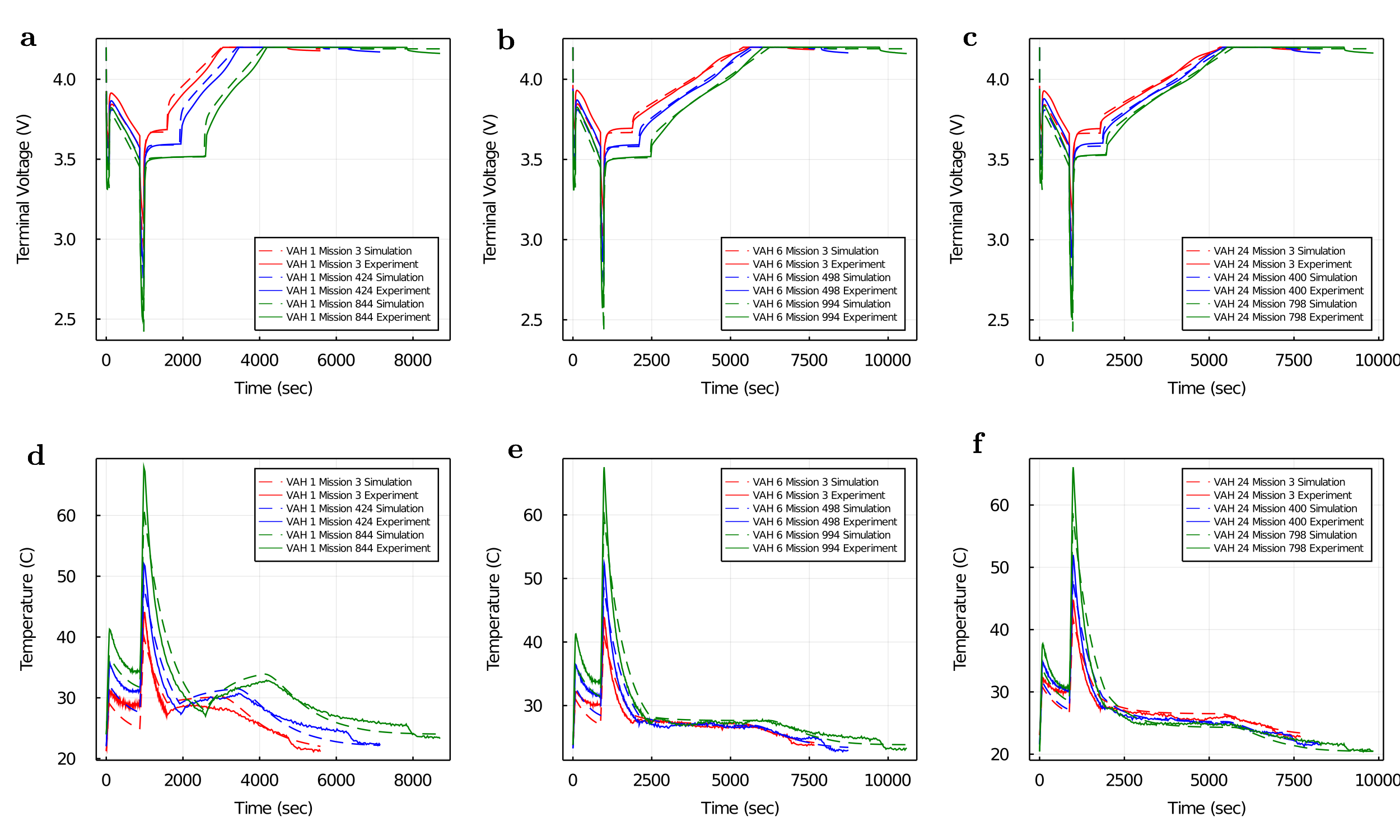}
        \caption{\textbf{\textit{Cellfit} voltage and temperature predictions} Voltage (a-c) and Temperature (d-f) predictions along with experimental data generated at a sub-milliecond run time of \textit{Cellfit} for 3 cells in the dataset. From left to right, the cells shown are serial numbers VAH01, VAH06, and VAH24, corresponding to two cells from the training set (VAH01 and VAH06) and one cell from the testing set (VAH24). The cells are each shown at a discharge cycle in the beginning, middle, and end of their lives. Importantly, the predictions are accurate for both temperature and voltage, even at the particularly critical for safety and difficult to predict end of landing high discharge point. Additionally, the performance predictions are nearly as good at the end of life as they are at the beginning of life.}
        
        \label{figure2}
\end{figure}

Cell voltage and temperature predictions from \textit{Cellfit} for six cells at different points in their lifetimes are shown in \cref{figure2}. We note that two of these missions (VAH01 and VAH06) are in the training set, meaning that the 1st cycle from each of these cells has been used for performance parameter estimation (aging parameter estimation is performed for all cycles here, and their predection will be discussed later). However, the later cycles have not been used in parameter estimation, meaning that the discharge curves shown have not been used directly for estimation of the performance parameters. As identified previously, particularly important for the application of eVTOL design, the model accurately predicts the highest temperature of the cell in all cases. The mean square error for temperature and voltage as well as the error of maximum temperature prediction for each cycle shown in \cref{table1}. Maximum temperature error stays below 8 degrees C in all cases, while mean square error of voltage stays well below .005 $V^2$ for all cases. It is interesting to note that MSE decreases with cycles (though the maximum temperature error increases), indicating that the model may actually grow slightly more accurate as the battery ages. Much of the thermal error is from the first rest period where the cell is cooling. This error can be attributed to the varying cooling conditions (convective cooling coefficient) experienced by the cell during the experiments. While we estimate the ambient temperature for each cycle, we do not consider variation in the cooling coefficient. This error could be mitigated by changing the cooling conditions in the simulation without loss of model generality (the cooling conditions can be considered external to the model).

\begin{table}
\npdecimalsign{.}
\nprounddigits{3}
\begin{tabular}{|l|r|n{2}{3}|n{2}{2}|n{2}{2}|}
\hline
Cell  & \multicolumn{1}{l|}{Cycle Number} & \multicolumn{1}{l|}{Voltage MSE} & \multicolumn{1}{l|}{Temp MSE} & \multicolumn{1}{l|}{Max T Error} \\ \hline
VAH01 & 3                                 & 0.004090612666                   & 5.908999584                   & 4.05691568                       \\ \hline
VAH01 & 424                               & 0.002427546384                   & 3.907546667                   & 3.492297486                      \\ \hline
VAH01 & 844                               & 0.002623913967                   & 4.295405545                   & 6.804419759                      \\ \hline
VAH06 & 3                                 & 0.003213887092                   & 1.98348633                    & 2.66814614                       \\ \hline
VAH06 & 498                               & 0.001746589809                   & 3.0471499                     & 3.729126767                      \\ \hline
VAH06 & 994                               & 0.001655835727                   & 4.39934389                    & 7.105003576                      \\ \hline
VAH24 & 3                                 & 0.003347372128                   & 2.260283797                   & 2.787835805                      \\ \hline
VAH24 & 400                               & 0.002100472814                   & 3.088887241                   & 3.801385092                      \\ \hline
VAH24 & 798                               & 0.001938728876                   & 4.638464608                   & 7.411882694                      \\ \hline
\end{tabular}
\caption{\label{table1}Voltage and Temperature MSE along with Maximum Temperature Error for the cycles shown in  \cref{figure2}}
\npnoround
\end{table}

\section*{Aging Modeling}
The discharge curves shown in \cref{figure2} are important to model throughout the life of the aircraft to inform design decisions based on changes in the battery performance characteristics over time and to ensure that the critical design case (when the battery is near end of life) can be properly accounted for in the design process. To model the changes in performance characteristics over the batteries life, the aging parameters must be allowed to evolve over the life of the battery. 

To estimate and model the aging parameters, those parameters must first be selected. This, in a process akin to hyperparameter optimization,\ssr{should we add a citation?} is accomplished by first picking a set of initial parameters (typically based on physical intuition), then calculating the loss function for each cycle of the cell. The aging parameters are the smallest set of parameters which produce a loss function which over the life of the cell is low and flat, indicating that the model, with estimated aging parameters and constant non-aging performance parameters, is accurate and remains accurate for the duration of the cell's life. An objective function, e.g. deviation from initial loss, could be used to evaluate whether the parameter set is sufficient to capture the variations.

\begin{figure}
    \centering
        \includegraphics[width=\textwidth]{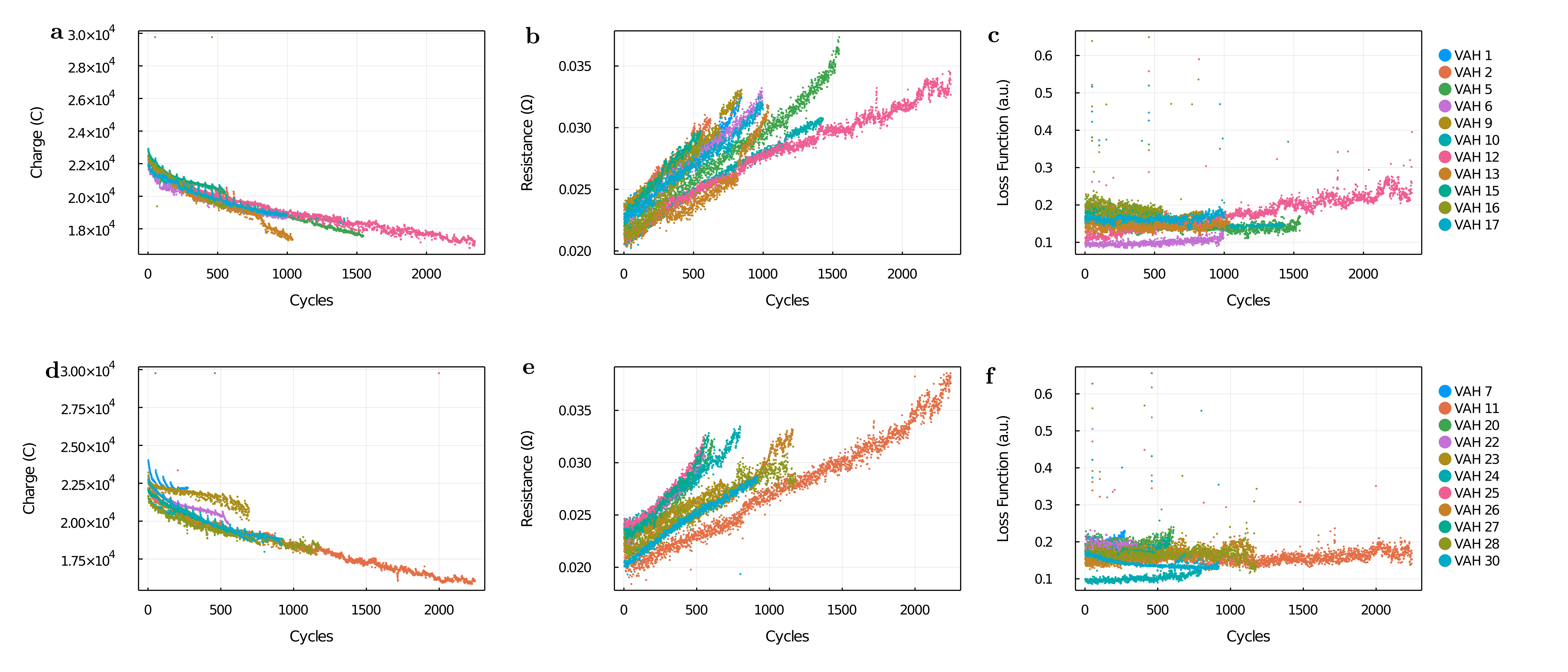}
        \caption{\textbf{Comparison of UBDM and Mechanistic Degradation Model}   Rows represent different cells, and columns represent different plots. The top row shows VAH01 from the training set, the middle row shows VAH24 from the interpolative test set, and the bottom row shows VAH07 from the extrapolative test set. From left to right, the columns show a comparison of resistance from the generated data and the UBDM and the MDM, a comparison of charge from the generated data and the UBDM and the MDM, and }
        \label{figure3}   
\end{figure}

\cref{figure3}c and f show the evolution of the loss function with cycle life for the training set. The performance parameters have been estimated using the first cycle from each cell. The aging parameters for each cycle are estimated directly using the grid-based approach. The loss function does not substantially increase for most cycles with increasing cycle number, and stays relatively low throughout the cycles. \cref{figure3}a and d show resistance for the training and testing set respectively, while \cref{figure3}b and e show total available charge for the the training and testing sets. Both resistance and charge show reasonable values for this cell (the charge parameter can be converted to a capacity by multiplying by the estimated filling fraction, which at the beginning of life results in a capacity of around 3mAh.) Additionally, they both follow established trends, with resistance following a roughly linear profile with different cells deviating to superlinear or sublinear processes, and charge being roughly proportional to the inverse square root of time near the begining of life, followed by a linear period \cite{Schmalstieg2014}. Once the aging parameters are fit to each cycle, a degradation model which governs the evolution of those parameters can be fit to the generated data. 

There are several advantages to the degradation modeling approach followed here. Firstly, the parameter estimation process for each cycle over the lifetime of the battery (the aging parameters) can be conducted independently, therefore, can be conducted in parallel, utilizing modern high performance computing architectures such as multi-core architectures and graphics processing units. Estimating the parameters in parallel theoretically allows us to estimate the parameters for each cycle over entire lifetime of a battery in the time required for a single cycle. Secondly, this approach results in accurate prediction of discharge characteristics throughout the battery's lifetime, which is vital for design, analysis, and safe operation of electric aircraft. Thirdly, the approach is model-agnostic, meaning that the degradation model can be physics-based, empirical, data driven, or any combination of them.

\section*{Degradation Modeling}
We develop two degradation models: a physics-based mechanistic degradation model (MDM), which includes contributions from three different degradation mechanisms: SEI growth, lithium plating, and active material loss\cite{Reniers2019,JIN2017750,YANG201728} and a novel approach, called the universal battery degradation model (UBDM), which is based on neural differential equations and universal ordinary differential equations\cite{chen2018neural,rackauckas2020universal}. In the UBDM approach, we use a neural network to supplement the mechanistic model to capture complex effects of charge and active material loss as well as change in resistance. Any mechanistic model can be substituted for the currently used mechanistic model in the UBDM, to account for changes in chemistry, operating conditions, cell models, etc. Additionally, any non-mechanistic model can be substituted for the simple neural network used in this work for the non-mechanistic model.

The MDM and UBDM are then trained alongside each other. For the estimation of these parameters, due to the high dimensionality and lack of interpretability of the neural network, we use a local optimization method (Adam)\cite{Kingma2015}. Derivatives are calculated using the julia differential equations suite, using the forward sensitivity analysis feature and the ForwardDiff.jl package\cite{rackauckas2017differentialequations,RevelsLubinPapamarkou2016}. The loss function for this estimation is the weighted sum of the mean square errors of resistance and charge. 

\begin{figure}
    \centering
        \includegraphics[width=\textwidth]{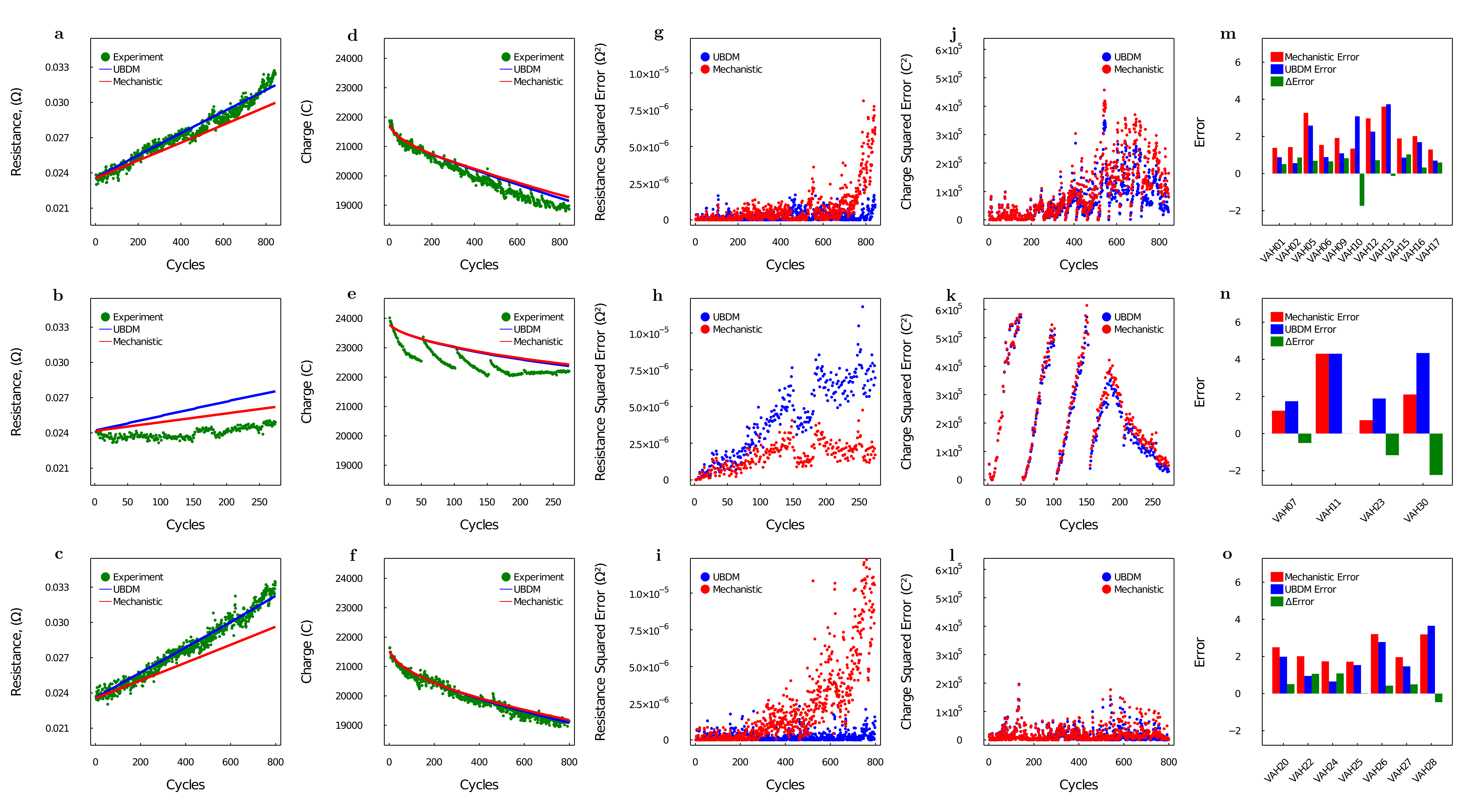}
        \caption{\textbf{Comparison of UBDM and Mechanistic Degradation Model}   From top to bottom, rows represent VAH01 from the training set, VAH07 from the Extrapolative Testing Set, and VAH24 from the Interpolative Testing Set, respectively (except for the last columns which show error from all cells in each set). From left to right, columns show resistance from the generated data, the UBDM, and the MDM, charge from the generated data, the UBDM, and the MDM, Resistance squared error from the UBDM and MDM, charge squared error from the UBDM, and the weighted sum of errors from all cells in each set.}
        \label{figure4}   
\end{figure}

The last column of \cref{figure4} shows the total error experienced by training and both test sets. For the training set and the interpolative test set, the UBDM outperforms the MDM. In the training set, the error of the UBDM shows a $19.18\%$ improvement over the MDM. In the interpolative testing set, the UBDM shows a $20.21\%$ improvement over the MDM. On the other hand, the UBDM performs worse than the MDM on the extrapolative testing set. The first four columns of \cref{figure4} provide some deeper insight on these results. The first row corresponds to VAH01, the baseline case. In this case, the resistance increases more than the MDM alone would indicate (\cref{figure4}a). The UBDM corrects this by increasing the rate of change of resistance. The same effect is seen in the second row with VAH24, from the interpolative test set (\cref{figure4}f). However, in the third row, which corresponds to VAH07 (the baseline case but only charged to 4.0V rather than 4.2V), the actual resistance grows less than the MDM would predict, yet the UBDM has not learned this bifurcation (\cref{figure4}k). The model was not trained on any cases in which the charging voltage was lower than 4.2V. In some cases, the UBDM is able to learn when resistance should be sublinear, such as in VAH12, which is available in the SI. More training data, or an accurate representation of all conditions which could be encountered by an eVTOL (essentially eliminating all possibility of extrapolation) would greatly increase the accuracy of the UBDM. The UBDM performs better on the charge estimation on the extrapolative test set. In all cases shown here, the UBDM outperforms the MDM on charge estimation, correcting it to account for lower charge. All training and test cases are available in the supporting information.

\begin{figure}
    \centering
        \includegraphics[width=\textwidth]{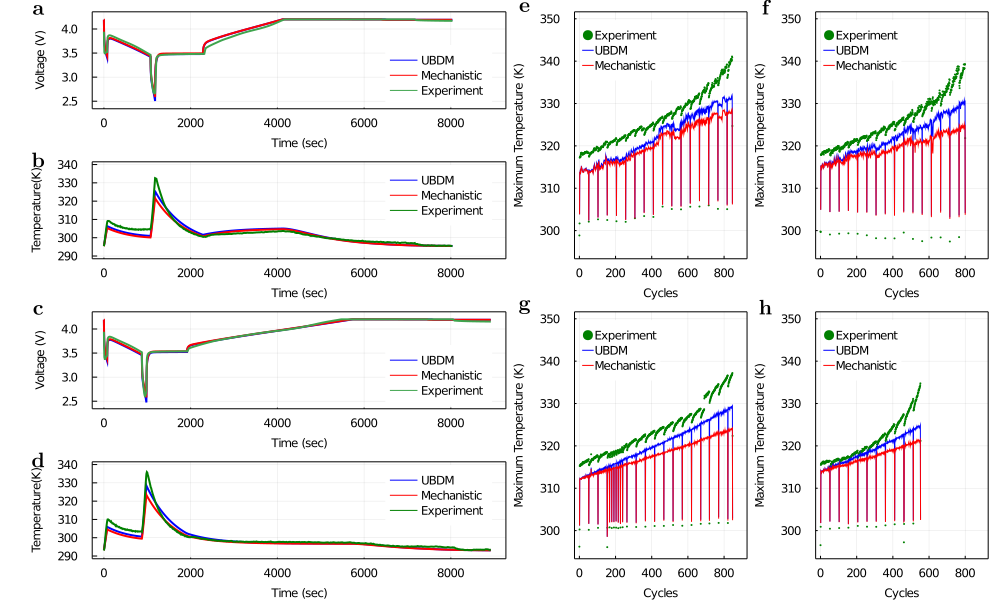}
        \caption{\textbf{Comparison of UBDM and MDM performance predictions}   \textbf{a-b} Voltage and Temperature predicitons for VAH15 cycle 500 and VAH24 cycle 750 \textbf{d-f} Maximum temperature predictions from the MDM and the UBDM for cells VAH01,VAH09, VAH24, and VAH07, respectively}
        \label{figure5}   
\end{figure}

While important, the model parameters of charge and resistance do not directly correspond to quantities used by aircraft designers. Therefore, it must be shown that improving the degradation model directly cascades to improvements of the performance model, particularly improvements in accuracy of predictions of voltage and temperature. \cref{figure5}a and b show voltage and temperature predictions for VAH15 and VAH24 (one from the training set and one from the interpolative testing set). While the UBDM does not seem to have a significant impact on the voltage predictions, it does have an impact on themperature predictions, particularly on the maximum temperature. In both cycle profiles shown here, the UBDM is more accurate than the MDM at predicting maximum temperature. This is expanded in \cref{figure5}d-f. In nearly all cases, the maximum temperature predictions are substantially improved by the UBDM over the MDM. Interestingly, this is true \textbf{even when the UBDM did not improve the overall accuracy as previously defined of the degradation model} For example, \cref{figure5}f shows VAH07, the same cell from the extrapolative test which was discussed previously. The maximum temperature predictions for VAH07 are improved by the UBDM. The most likely explanation for this phenomenon is the improvement of the charge predictions of the UBDM for this cell. Indeed, as can be seen in the SI, the UBDM shows an improvement in maximum temperature predictions for all but 4 of the cells as compared to the MDM.

\section*{Conclusions}
In this work, we have generated a battery performance and thermal response dataset specific to the urban air mobility use-case.  We expect this dataset to rapidly accelerate the nascent field of batteries for electric aircraft.  In order to learn this dataset, we develop a `universal' battery modeling approach, which combines exact physical laws along with neural networks that can learn the residual (missing) physics. Using this modeling approach, we show incredible accuracy and computational efficiency for voltage, temperature and degradation for the dataset. This high accuracy and computational efficiency, along with the flexible nature of the UBDM should provide a valuable tool for eVTOL designers, infrastructure designers, and others who need a fast and accurate tool for battery modeling.


\begin{methods}
\subsection{Cell Testing and Data Generation:} 
This work utilized Sony-Murata 18650 VTC-6 cells (INR technology). This cell has a rated capacity of 3000mAh at a nominal voltage of 3.6V. The manufacturer specified maximum continuous discharge rate is 10C with a 80$^\circ$C upper temperature cut off. This cell is appropriate for evaluation in eVTOL applications as it can sustain high power demand while providing a cell specific energy of 230 Wh/kg.

All cells were tested in a Arbin 200A cylindrical cell holder paired with a BioLogic BCS-815 modular battery cycler. As shown in Supplementary Figure 1, cell testing was performed in enclosures that were placed in a temperature chamber that was maintained at 25$^\circ$C. Cell can surface temperatures were measured via a thermocouple fixed to the cell body with aluminum tape.

\begin{table}
\begin{center}
    \begin{tabular}{||c c c c c||}
        \hline
        Phase & Duration & Power & C-rate & Discharge Energy \\
        \hline\hline
        Take-off & 75s & 54W & 5C & 1.12Wh \\
        \hline
        Cruise & 800s & 16W & 1.48C & 3.55Wh \\
        \hline
        Landing & 105s & 54W & 5C & 1.57Wh \\
        \hline
   \end{tabular}
   \caption{Baseline Mission Parameters (Discharge)}
  \label{baseline}
\end{center}
\end{table}

\begin{table}
\begin{center}
    \begin{tabular}{||c c c||}
        \hline
        Phase & Definition & End Criteria\\
        \hline
        Rest 1 & 0 Amps & $T< 27^{\circ}C$\\
        \hline
        CC Charge & 1C & $V>4.2$\\
        \hline
        CV Charge & V=4.2 &$I<C/30$ \\
        \hline
        Rest 2 & I=0 &$ T<35^{\circ}C$ \\
        \hline
   \end{tabular}
   \caption{Baseline Mission Parameters (Charge)}
  \label{baselinecharge}
\end{center}
\end{table}

Generation of the eVTOL data set was developed from the baseline power profile for a notional eVTOL flight given in table \ref{baseline}. Battery cell state-of-health was measured for the baseline mission profile and a series of variations on the baseline mission. The variations, as shown in \cref{figure1}, included modifying a single variable in the baseline mission profile. The test descriptions are given in \cref{baselineVarTrain} and \cref{baselineVarTest}. For each mission profile, cells were cycled until cell voltage reached 2.5V or cell temperature reached 70$^\circ$C during discharge. To measure cell energy capacity a reference performance test was performed at the start of each aging test campaign and following each subsequent set of 50 mission cycles. Each RPT consisted of measuring capacity by discharging the cell from 100\% to 0\% state-of-charge at a discharge rate of C/5 until the voltage dropped below 2.5V and an internal resistance measurement at 20\% and 60\% depth of discharge. After all discharge cycles a rest time was observed to allow the cell to cool to 30$^\circ$C. Once 30$^\circ$C cell temperature was reached a CC-CV charge profile was started with constant current charge at 1C, followed by constant voltage charge until charge current decayed to C/30 to a (nominally) 4.2V end of charge voltage. 

\begin{table}
\centering
\begin{tabular}{||c|l||}
        \hline
        \multicolumn{2}{||c||}{Training Data} \\
        \hline
        Mission Profile & Description \\
        \hline
        VAH01 & Baseline mission \\
        VAH02 & Cruise time, $t_c =1000$s \\
        VAH05 & Power reduction of 10 percent for takeoff, cruise, and landing \\
        VAH06 & CC-CV charge cycle at C/2 CC – C/30 CV \\
        VAH09 & Thermal chamber temperature of 20$^\circ$C\\
        VAH10 & Thermal chamber temperature of 30$^\circ$C \\
        VAH11 & Power reduction of 20\% for takeoff, cruise, and landing \\
        VAH12 & Cruise time, $t_c =400$s \\
        VAH13 & Cruise time, $t_c =600$s \\
        VAH15 & Cruise time, $t_c =1000$s \\
        VAH16 & CC-CV charge cycle at 1.5C CC – C/30 CV \\
        VAH17 & Baseline mission \\
         \hline
     \end{tabular}
\caption{Descriptions of the Training dataset}
\label{baselineVarTrain}
\end{table}

\begin{table}
\centering
    \begin{tabular}{||c|l||}
        \hline
        \multicolumn{2}{||c||}{Testing and Validation Data} \\
        \hline
        Mission Profile & Description \\
        \hline
        VAH07 & End of charge voltage at 4.0V \\
        VAH20 & CC-CV charge cycle at 1.5C CC – C/30 CV \\
        VAH22 & Cruise time of 1000s \\
        VAH23 & End of charge voltage at 4.1V \\
        VAH24 & CC-CV charge cycle at C/2 CC – C/30 CV\\
        VAH25 & Thermal chamber temperature of 20C \\
        VAH26 & Cruise time of 600s \\
        VAH27 & Baseline \\
        VAH28 & Power reduction of 10 percent for takeoff, cruise, and landing \\
        VAH30 & Thermal chamber temperature of 35$^\circ$C \\
        \hline
   \end{tabular}
\caption{Descriptions of the Testing and Validation dataset}
\label{baselineVarTest}
\end{table}

\subsection{Data cleanup and handling:}
\textbf{-- (i) rest times for the 3 Test missions (ii) mission cycles/ phantom cycles -- how we handle them}

\subsection{Cell Model:}

The cell model, adapted from one presented by Daigle and Kulkarni \cite{Daigle2013} is a simplified coupled electrochemical model consisting of a system of ordinary differential equations, allowing it to be much faster than the coupled system of algebraic and partial differential equations of more complex battery models. It has two principal components: Modeling charge (q) movement and potential (V). The model comprises 4 domains: bulk and surface regions of both the cathode and the anode. Charge flow is governed by applied current between the surfaces of each electrode and by diffusion between the surface and bulk of each electrode, as described in equations \eqref{Diffusion}, where the subscript $b$ represents bulk, $s$ reprents surface, $p$ represents positive, and $n$ represents negative $D$ represents a diffusion coefficient.  Note that an additional charge transfer term will be added to the model to represent degradation via charge loss later. These equations conserve charge, allowing this model to be faithful to the physics of a battery.

\begin{subequations}
\begin{equation}
    \mathrm{\frac{dq_{sp}}{dt} = \frac{1}{D}(\frac{q_{bp}}{v_{bp}}-\frac{q_{sp}}{v_{sp}})+I_{app}}
\end{equation}
\begin{equation}
    \mathrm{\frac{dq_{bp}}{dt} = -\frac{1}{D}(\frac{q_{bp}}{v_{bp}}-\frac{q_{sp}}{v_{sp}})}
\end{equation}
\begin{equation}
    \mathrm{\frac{dq_{sn}}{dt} = \frac{1}{D}(\frac{q_{bn}}{v_{bn}}-\frac{q_{sn}}{v_{sn}})-I_{app}}
\end{equation}
\begin{equation}
    \mathrm{\frac{dq_{bn}}{dt} = -\frac{1}{D}(\frac{q_{bn}}{v_{bn}}-\frac{q_{sn}}{v_{sn}})}
\end{equation}
\label{Diffusion}
\end{subequations}

Voltage is calculated using a buildup consisting of 5 terms: cathode and anode equilibrium potentials, cathodic and anodic surface overpotentials, and an ohmic overpotential. All overpotentials are also passed through first-order filters to prevent sudden voltage changes. Surface kinetics of each electrode are governed by the Butler-Volmer equations, as shown in \eqref{surfaceBV} where $J_{p,n}$ is the current density, $J_{p0,n0}$ is the exchange current density, $V$ is the overpotential, and $\tau$ is the time constant for each overpotential.  

\begin{equation}
    \mathrm{\frac{dV_{\eta~p, n}}{dt} = \frac{\frac{RT}{F\alpha}asinh(\frac{J_{p,n}}{2J_{p0,n0}})-V_{\eta p,n}}{\tau_{\eta~p,n}}}
    \label{surfaceBV}
\end{equation}

Cell resistance is modeled by a lumped resistance parameter, and the ohmic overpotential is given in equation \eqref{ohmic} .  

\begin{equation}
    \mathrm{\frac{dV_{ohm}}{dt} = \frac{IR-V_{ohm}}{\tau_{ohm}}}
    \label{ohmic}
\end{equation}

The anode and cathode equilibrium potentials are modeled by the Redlich-Kister polynomials, and are given in equation \eqref{equilibrium}, where $V_{U n,p}$ is the reference potential, $R$ is the ideal gas constant, $n$ is the number of transferred electrons, $F$ is Faraday's constant, $x_{n,p}$ is the positive and negative filling fraction (respectively), and $A_{n,p}$ are the fitting coefficients.\cite{Karthikeyan2008} Importantly, while the fitting coefficients are estimated using simulated annealing, they are constrained to obey the second law of thermodynamics. This constraint helps to prevent overfitting and increases the interpretability of the model. It is enforced within the simulated annealing routine, by not allowing the routine to output an OCV curve which is not monotonic.

\begin{equation}
    \mathrm{V_{U n,p} = V_{U0 n,p} + \frac{RT}{nF} \log\Big(\frac{1-x_{n,p}}{x_{n,p}}\Big)+\frac{1}{nF}\sum_{i=1}^{N_{n,p}} A_{n,p}(2x_{n,p}-1)^{i} - \frac{2x_{n,p}(i-1)(1-x_{n,p})}{(2x_{n,p}-1)^{1-i}}}
    \label{equilibrium}
\end{equation}

The thermal model is based on a lumped parameter model with heat generation terms for ohmic heating and entropic heating. Cooling is convective cooling, owing to the test setup as described above. The equations for the thermal model is given in equation \eqref{thermal}, where $I$ is the current, $R$ is the resistance, $\frac{\partial U}{ \partial T}$ is the entropic coefficient, $h$ is the conductive coefficient, $A$ is the surface area (because the surface area of all cells is the same, we model $hA$ as a lumped parameter), $m$ is the cell mass, and $c_{p}$ is the specific heat.

\begin{equation}
    \mathrm{\frac{dT}{dt} = \frac{I^{2}R+\frac{I\frac{\partial U}{ \partial T}}{nF} - hA(T-T_{amb})}{mc_{p}}}
    \label{thermal}
\end{equation}

The cooling model is specific to this dataset; any cooling model can be used in lieu of newtonian convection to best approximate cooling for any situation.

\subsection{Parameter Estimation and Identification}

Different parameters were estimated using different techniques. Performance parameters, including all Redlich-Kister coefficients, the diffusion coefficient, the reference potentials, all thermal parameters, and time constants were estimated using simulated annealing. A stochastic optimization technique was chosen to avoid the parameter identifiability issues described previously. The simulated annealing algorithm begins by evaluating the loss function at an initial point $\theta$. Then, using gibbs sampling, a random parameter is chosen to iterate, and a new parameter vector $\theta_{candidate}$ is created by adding a step $\lambda\mathcal{N}(0,1)$ where $\lambda$ depends on the magnitude of the selected parameter, and the loss function is then evaluated at $\theta_{candidate}$. If the new loss function is lower than the previous loss function, then $\theta$ is replaced by $\theta_{candidate}$. If the new loss function is higher than the previous loss function, then $\theta$ is replaced by $\theta_{candidate}$ with a probability calculated by the Boltzmann distribution:

\begin{equation}
    \mathrm{P_{acceptance} = e^\frac{(L(\theta_{candidate})-L(\theta))}{T}}
    \label{annealprobability}
\end{equation}

Over the course of the optimization, the temperature $T$ is lowered to reduce the likelihood of moving to a less optimal parameter vector. This enables the algorithm to exploit and explore the parameter space. Exploration occurs when the temperature is relatively high, because the algorithm will allow movement to a higher loss value. This allows the algorithm to escape some local minima. Exploitation occurs when the temperature is relatively low, forcing the algorithm to move towards the local minima.

We identified the aging parameters (charge and resistance) using a grid-based method. We chose this method for two reasons. Firstly, because there were only two parameters, and because of the aforementioned speed of the calculation of the loss function for this problem, a grid based method was tractable. Secondly, there was relatively little tolerance for optimization noise in these parameters, and the grid based method produced much less noise than a stochastic method such as simulated annealing. In the grid based method, upper and lower limits for charge and resistance were first identified based on physical constants. The lower resistance limit was defined as 0.01 ohms, and the upper resistance limit was defined as 0.05 ohms. The upper charge limit was defined as 26000 coulombs, and the lower charge limit was set to around 15000 coulombs. Using these limits, a 10 by 10 grid was created and the loss function was evaluated at each point on the grid. At the best points (those with the lowest loss function) the grid was further subdivided and the loss function was re-evaluated for a specified number of subdivisions and number of subgrid points.

Identifying the degradation parameters, including both the mechanistic and UBDM degradation parameters, involved a two-stage process. In the first stage, the parameters were initialized using simulated annealing following from a random starting point, as described above. In the second stage, the parameters were refined using a gradient-based optimization algorithm, Adam. This two-stage process was chosen to solve two problems: firstly, when integrated for a long time interval, such as a battery lifetime, random initialization of Neural Network parameters as is the norm in most machine learning applications can results in highly unstable systems, resulting in unphysical concentrations due to large charge losses or negative charge losses (charge gains). It is often not possible to calculate a gradient based on simulations with unphysical results which cause floating point errors, so a stochastic optimization method is required. Secondly, it is easy to see that initializing a gradient based optimizer with the zero vector results in an equilibrium point at the initialization point, and therefore, training does not occur. Additionally, there are too many parameters in the UBDM to ensure convergence of simulated annealing, and to give the two algorithms a fair comparison requires giving them the same training protocols.

As briefly mentioned in the article, we used a custom loss function to identify the parameters. That function given in equation \eqref{lossfunc}, where a tilde over the variable indicates that it is the average experimental of the data, and a t subscript indicates that it is taken at timestep t, $W$ indicates the weight given to each part, and the subscript of the $W$'s indicates the part of the loss function to which that weight is applied. In this work, the weight was 10 for the voltage error and 1 for the other error terms. 

\begin{equation}
    \mathrm{L = \sum_{t} (W_{V}\frac{|V_{t}-\Tilde{V_{t}}|}{\Bar{V}}+W_{T}\frac{(T_{t}-\Tilde{T_{t}})^{2}}{\Bar{T}})+W_{MT}\frac{|max(T_{t})-max(\Tilde{T_{t}})|}{\Bar{T}}}
    \label{lossfunc}
\end{equation}

\textbf{Degradation Modeling}
 To model battery degradation, we follow a multistep process with three steps: first, find the parameters which change with aging. Then, fit those parameters to each (SOC) cycle for any given battery life-cycle, and identify whether or not the loss function is increasing as the battery ages \cite{Daigle2016}. If the loss function is not increasing with time, then the parameters identified are the correct set of aging parameters. If the loss function is increasing with time, then the selection of the aging parameters needs to be revisited. As mentioned previously, because of the requirement for relatively low amounts of noise and the relatively low computational cost of estimating these parameters, the aging parameters were estimated using a grid-based method\cite{KURCHIN2019161}.

Using this methodology, the precise degradation model can be chosen to fit the generated data once the parameters for each cycle have been estimated. For this work, we employed two separate degradation models. The first we refer to as the mechanistic model. It is based on physical principles and consists of charge loss due to SEI formation, \cref{seigrowth}, active material isolation, \cref{amisolation}, and lithium plating, \cref{plating}\cite{JIN2017750, YANG201728}. Additionally, we add a resistance increase term, Eq. \eqref{resistancegrowth}\cite{Daigle2016}.
\begin{subequations}
\begin{equation}
    \mathrm{\frac{dQ_{sei}}{dt} = \frac{K_{SEI}e^{\frac{-E_{SEI}}{RT}}}{2(1+\lambda\theta)\sqrt{t}}}
    \label{seigrowth}
\end{equation}
\begin{equation}
    \mathrm{j_{pl} = i_{0pl}e^{-\frac{0.5F}{RT}(V_{Un}-V_{\eta n})}}
    \label{plating}
\end{equation}
\begin{equation}
    \mathrm{\frac{dQ_{am}}{dt} = K_{AM}e^{-\frac{E_{AM}}{RT}}SOCI_{app}}
    \label{amisolation}
\end{equation}
\begin{equation}
    \mathrm{\frac{dR}{dt} = w_{d}|I_{app}|}
    \label{resistancegrowth}
\end{equation}
\end{subequations}

We additionally have proposed a novel degradation model which uses Universal Ordinary differential Equations (U-ODE) to estimate charge loss and resistance change. Neural Ordinary differential equations (NODE) are a model where the derivative of a function is approximated using a neural network\cite{chen2018neural}. U-ODE's are an extention of NODE's where only part of the time derivative of the system is approximated using a neural network\cite{rackauckas2020universal}. In this case, we have supplemented our mechanistic degradation model with a U-ODE to improve accuracy and generalization of predictions. The U-ODE neural network is a function of all of the other state variables of the system, as shown in \eqref{uode}. The full degradation model is given in equations \eqref{fulldegradation}

\begin{subequations}
\begin{equation}
    \mathrm{\frac{dQ_{nonmechanistic}}{dt} = NN(u,\theta)}
\end{equation}
\begin{equation}
    \mathrm{\frac{dR_{nonmechanisitic}}{dt} = NN(u,\theta)}
\end{equation}
\label{uode}
\end{subequations}
where $u$ is defined as 
\begin{equation}
    \mathrm{u = [q_{sp},q_{bp},q_{sn},q_{bn},V0,V_{\eta n},V_{\eta p},T,R,q_{max}]^{T}}
\end{equation}
and $\theta$ are the parameters of the neural network. We combine the mechanistic and non-mechanistic parts of the model to arrive at our final degradation model, the UBDM, which is written as:
\begin{equation}
    \mathrm{\frac{dQ}{dt} = \frac{dQ_{SEI}}{dt}+\frac{dQ_{AM}}{dt}+j_{pl}+\frac{dQ_{nonmechanistic}}{dt}}
    \label{fulldegradation}
\end{equation}

\begin{equation}
    \mathrm{\frac{dR}{dt} = \frac{dR_{mechanistic}}{dt}+\frac{dR_{nonmechanistic}}{dt}}
    \label{fullresistance}
\end{equation}

The loss of charge in the degradation model always occurs at the anode surface, meaning that it consists of a subtraction term added to $q_{sn}$, which is consistent with most research on degradation occurring at the anode. 



\end{methods}


\subsection{References}
\bibliography{References.bib}


\begin{addendum}
\item This work was supported by Airbus A$^{\wedge}$3. The authors would also like to thank Ananth Sridharan for his inputs on the relevance of battery SOC and SOH estimation for electric aircraft sizing, and Romain Teulier for coordinating the battery testing presented in the article.  E. F. and D. C. performed the work reported in this paper during their time as employees at Airbus A$^{\wedge}$3.

\item[Contributions] A.B. and L.F. were the main developers of the software code for Cellfit.  S.S., A.B., L.F. and V.V. developed the underlying modeling approach.  M.G. and V.V. developed the simulated annealing optimization approach.  E.F. and D.C. coordinated generation of the experimental dataset.  A.B., S.S., E.F. and V.V. wrote the paper.  All authors read the paper and provided input.
\item[Competing Interests] The authors declare that they have no competing financial interests.
\item[Correspondence] Correspondence and requests for materials should be addressed to V. Viswanathan (email: venkvis@cmu.edu).
\end{addendum}

\clearpage
\includepdf[pages=-]{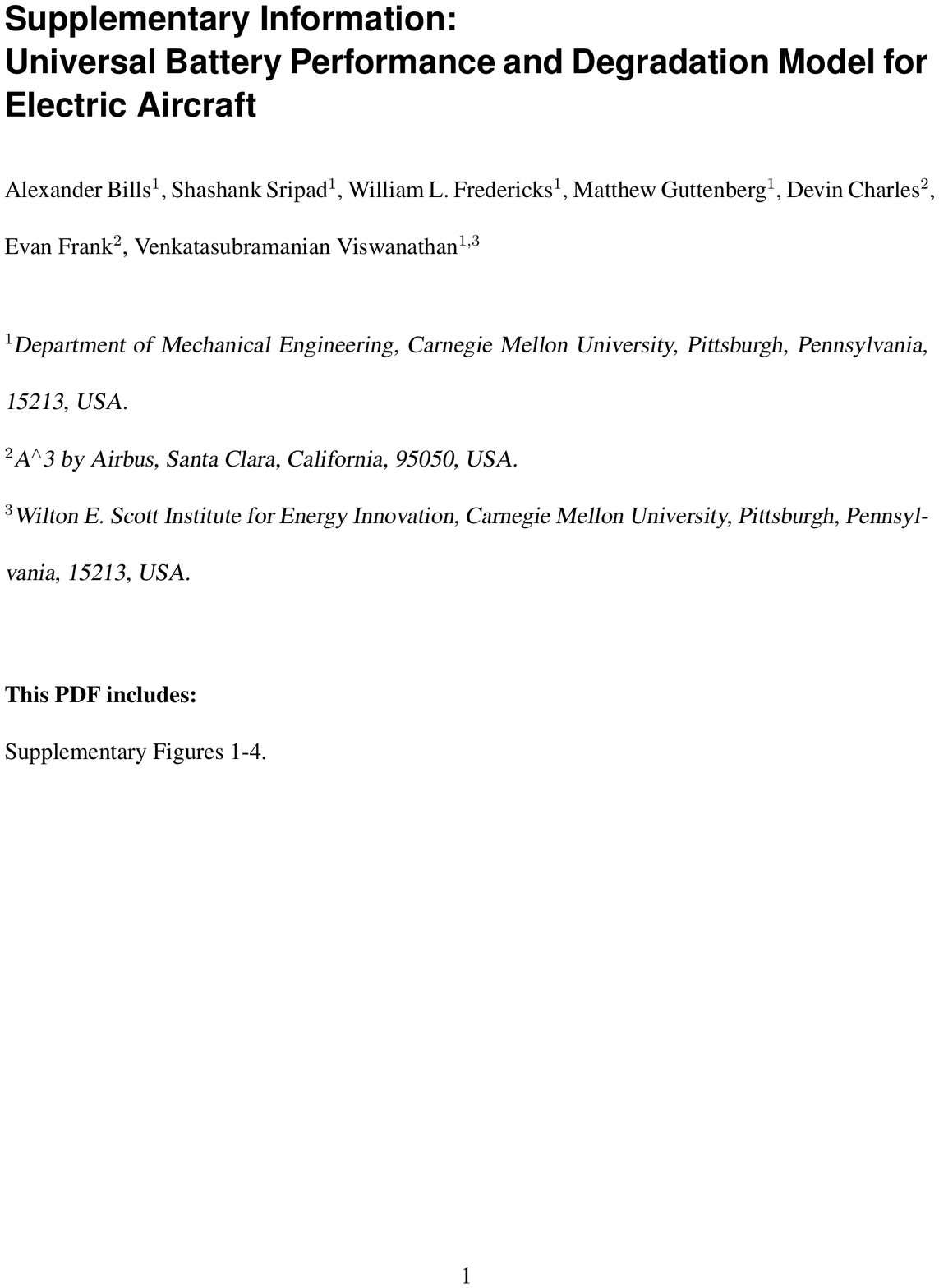}

\end{document}